%% file: hybrid_draft.tex
\documentclass[onecolumn, prl,superscriptaddress]{revtex4-2}
\usepackage{fullpage}
\usepackage{amsmath}
\usepackage{xcolor}
\usepackage{float}
\usepackage{graphicx}
\usepackage[parfill]{parskip}

\begin{document}

\title{Anderson Impurity Mechanism for a Multi-Level Model in $\delta$-Pu}
\author{Alexander R. Mu\~noz }
\affiliation{T-1: Physics and Chemistry of Materials; Los Alamos National Laboratory, Los Alamos NM 87544}
\author{Travis E. Jones}
\affiliation{T-1: Physics and Chemistry of Materials; Los Alamos National Laboratory, Los Alamos NM 87544}

\begin{abstract}
Electronic correlations and spin-orbit interactions in plutonium create variations in the bonding behavior of each of its allotropes.
In $\delta$-Pu, the 5f electrons lie at the tipping point between itinerant and localized behavior which makes the creation of predictive models very difficult.
We perform density functional theory calculations to study the effect of correlated descriptions on the mechanical properties of $\delta$-Pu.
We find that 7.5\% $E_{xx}$ in the HSE functional yields the experimental lattice parameters, moreover, this functional recovers the experimental elastic constants while other approximations fail.
The electronic structure of the hybrid functional yields several signatures of strong correlations including orbital-selective bonding of a single 5f electron and a pseudogap-like feature which work in tandem to improve the description of mechanical properties.
We show how the emergence of orbital-selective bonding in the hybrid functional can be understood through an Anderson impurity picture which predicts the augmentation of $\pi$-bonding, the decrease of $\sigma$-bonding, and expands the volume of the cell, enabling the accurate description of the mechanical properties of $\delta$-Pu. 
\end{abstract}

\maketitle

The understanding of 5f electron materials has made significant progress, but the presence of electronic correlations and spin-orbit coupling continue to make the actinides difficult to model. \cite{soderlind_2019}
Understanding and modeling the bonding of the 5f electrons in Pu is particularly challenging.
The 5f electrons in Pu lie at the intersection of the itinerant behavior of the early actinides and the localized behavior of the late actinides. \cite{albers_2001,hecker_2000}
As the most technologically important phase of Pu, the face-centered cubic $\delta$ phase has merited the most study over the years.
This has inspired partially localized, delocalized, and fluctuating models of the electronic structure. \cite{huang_2020,penicaud_1997,eriksson_1999,soderlind_2003,svane_2007}
Without the use of a mixed-level model, density functional theory (DFT) fails to correctly capture the volume of $\delta$-Pu.
The difficulty of describing $\delta$-Pu comes from its nearly localized electronic behavior requiring the partitioning of itinerant and localized electrons. \cite{soderlind_2010,xu_2008,boring_2000} 
Mixed-level models have produced accurate descriptios of $\delta$-Pu, but there is not a mechanism in place to validate their use.\cite{soderlind_2019,clark_2019,amadon_2016,soderlind_1997}

To date, it remains unclear if a hybrid density functional aids in the description of $\delta$-Pu, despite dynamical mean field theory and GW calculations indicating a need for electronic correlations. \cite{amadon_2016,tutchton_2020,kutepov_2023}
The inclusion of electronic correlations in $\delta$-Pu yields accurate descriptions of experimental photoelectron spectra and the Fermi surface. \cite{tutchton_2020}
A Hubbard U has been used in DFT studies, but hybrid functionals have shown large lattice expansions when including Hartree-Fock exact exchange, $E_{xx}$, in excess of 25\% in the PBE functional. \cite{shick_2005,raymond_2009}
However, the fraction of $E_{xx}$ is known to scale with the dielectric constant, and no studies have been performed on the effects of small percentages of $E_{xx}$ on any allotrope of Pu.
In fact, the literature has shown that using small $E_{xx}$ mixing has improved the description of many correlated systems. \cite{shimazaki_2015,skone_2014}

In this manuscript, we use the 3Q magnetic structure to study the effect of the Heyd-Scuseria-Ernzerhof (HSE) functional on $\delta$-Pu.\cite{heyd_2003}
The estimate of the dielectric constant in a metal is difficult, so we tune the exact exchange to the lattice constant to estimate the electronic screening and identify HSE with 7.5\% $E_{xx}$ mixing (HSE(7.5\%)) as a well-optimized functional for describing the mechanical properties of $\delta$-Pu.
We find that HSE(7.5\%) recovers the measured single crystal elastic moduli by allowing one 5f electron to participate in bonding, much like the mixed-level model.
By examining the real space charge density and the density of states, we uncover the mechanism of partial 5f localization in HSE(7.5\%).
The $f_{xyz}$ orbital in HSE(7.5\%) participates in $\pi$-bonding enabled by Anderson-like hybridization.
The symmetry of the bonding allows for the softening of the elastic modes.
The model from intersite correlations is validated by comparison to on-site interactions that selectively drive stiffening and softening of elastic constants through the creation of $\sigma$ and $\pi$ bonding, respectively.

We use DFT in the projector augmented wave method as implemented in the Vienna Ab initio Simulation Package (VASP). \cite{blochl_1994,kresse_1996,kresse_1999}
Calculations were done using 32 atom cells of $\delta$-Pu with first-order Methfessel-Paxton smearing with a width of 0.2 eV and a $6\times6\times6$ Monkhorst-Pack \textbf{k}-point mesh. \cite{methfessel_1989,monkhorst_1976}
The cutoff energy for the plane wave basis set was 500 eV, which is standard for $\delta$-Pu. \cite{soderlind_2019}
Spin-orbit coupling was included in all calculations.
The magnetic state used in our study is the noncollinear 3Q magnetic state that has been used to study the mechanical properties of $\delta$-Pu.\cite{rudin_2022,soderlind_2023}
This state is effective in describing $\delta$-Pu due to its proximity to the experimentally observed valence fluctuating state.\cite{lashley_2005,janoschek_2015,soderlind_2023}

Starting with the experimental lattice constant, $a_0$, we increment the lattice constant in steps of 0.01 $\text{\AA}$ to identify the equilibrium lattice constants, $a_{eq.}$, in DFT.\cite{ellinger_1956}  
The elastic constant calculations were computed using a four atom cell with a $12\times12\times12$ \textbf{k}-point mesh.
We strain the equilibrium lattice with volume-conserving strains to compute $C'$ and $C_{44}$.\cite{bercegeay_2005}
The volume-conserving strains used in this study strain the lattice by -0.5\%, -0.25\%, 0.25\% and 0.5\%.
We validate our results by comparing to the internal routines in VASP.

The functionals used in this study are the generalized gradient approximation of Perdew et al. (PBE), OFR2, PBE+U+J, PBE+U$_{\text{eff}}$, OFR2+U$_{\text{eff}}$, and the Heyd-Scuseria-Ernzerhof (HSE) functional with a variable fraction of $E_{xx}$. \cite{perdew_1996,kaplan_2022,dudarev_1998,liechtenstein_1995,heyd_2003}
While successful, the PBE functional introduces errors when describing systems with strong electronic correlations by delocalizing electrons and, in the case of $\delta$-Pu, PBE delocalizes the f-electrons leading to the contraction of the lattice constant. \cite{vega_2018,janthon_2013,gorni_2021}
The Hubbard U and Hund's J encode electronic correlations in the Hamiltonian.
We use U and J as parameters to fit the lattice parameter of $\delta$-Pu, Supplementary Material.
The HSE functional has an exchange-correlation term of the form,
\begin{equation}
  E_{xc}^{\omega\text{PBEh}} = \alpha E_x^{\text{HF,SR}}(\omega)+(1-\alpha)E_x^{\text{PBE,SR}}(\omega) +\\E_x^{\text{PBE,LR}}(\omega)+E_c^{\text{PBE}},
\end{equation}
where \text{LR} denotes long-range components of the exchange energy, $\text{SR}$ denotes the short-range components of the exchange energy, $\omega$ is a parameter controlling the range of the short-range interactions, and $\alpha$ is the mixing coefficient of the exact exchange from Hartree-Fock. \cite{heyd_2003}
We use $\alpha$ as a parameter to introduce interactions and accurately capture the screening in the system. 
We refer to individual HSE functionals as HSE($\alpha$\%).

To identify potential functionals, we need to introduce functionals that correctly describe the screening in $\delta$-Pu.
It is known that using the PBE functional underestimates $a_0$, and by increasing correlations, we increase the lattice parameter.
To identify the proper proportion of screening, we compute the cohesive energy with HSE(0\%), HSE(5\%), HSE(7.5\%), and HSE(10\%) for a set of lattice parameters, Supplement.
We followed the same approach to identify optimal values of U and J, Table 1.
As the $E_{xx}$ is increased in HSE, the $a_{eq}$ increases and the cohesive energy decreases in magnitude.
When $\alpha=7.5\%$, HSE is within one percent of $a_0$ while 10\% overestimates experiment and 0\%, 5\% and PBE underestimate the experimental lattice parameter. 
Despite Pu's metallic character, $\alpha=7.5\%$ is reasonable given the proposed pseudogap behavior in the $\delta$-phase.\cite{wartenbe_2022}

The optimal electronic screening is set by $a_0$, so we test the functional on another ground-state property, the elastic moduli.
The experimental and theoretical computed elastic properties are shown in Table~\ref{table:elastic}.
PBE decreases $a_{eq}$ and stiffens the elastic constants and the bulk modulus,
\begin{equation}
  B = \frac{C_{11}+2C_{12}}{3}.
\end{equation}
The bulk modulus is overestimated by 72 percent and $C_{11}$ is overestimated by 67 percent.
For HSE(7.5\%), we show the closest correspondence with experiment with a RMSE of 5.73 across C$_{11}$, C$_{12}$, and C$_{44}$.
This is in contrast to the RMSE of 32.3 for PBE.
DFT+U+J improves the bulk modulus because the lattice is expanded, but it does worse for $C_{44}$ indicating that on-site correlations are insufficient.
The HSE(7.5\%) results are the only results within errors of experiment.
The task is now to explain the differences in bonding that lead to the elastic behavior of PBE and HSE in their equilibrium volumes.

\begin{table}
    \centering
	\begin{tabular}{l|c|c|c|c|}
		Functional & B (GPa) & C$_{11}$ (GPa) & C$_{12}$ (GPa) & C$_{44}$ (GPa)  \\
		\hline
        Exp. & 29.9 & 36.3 & 26.7 & 33.6 \\
		PBE & 51.5 & 60.5 & 47.0 & 40.3 \\
		DFT+U+J & 40.1 & 53.1 & 33.6 & 45.5 \\
		HSE(7.5\%) & 26.7 & 37.8 & 21.3 & 34.8 \\
	\end{tabular} 
    \caption{Experimental and theoretical results for elastic moduli. All results are reported at $a_{eq}$. The DFT+U+J results are for the U = 1 eV and J = 0.3 eV case that coincides with $a_0$. Other cases of U and J are shown in the Supplementary Material.}
    \label{table:elastic}
\end{table}

\begin{figure}
\centering
\includegraphics[width=3in]{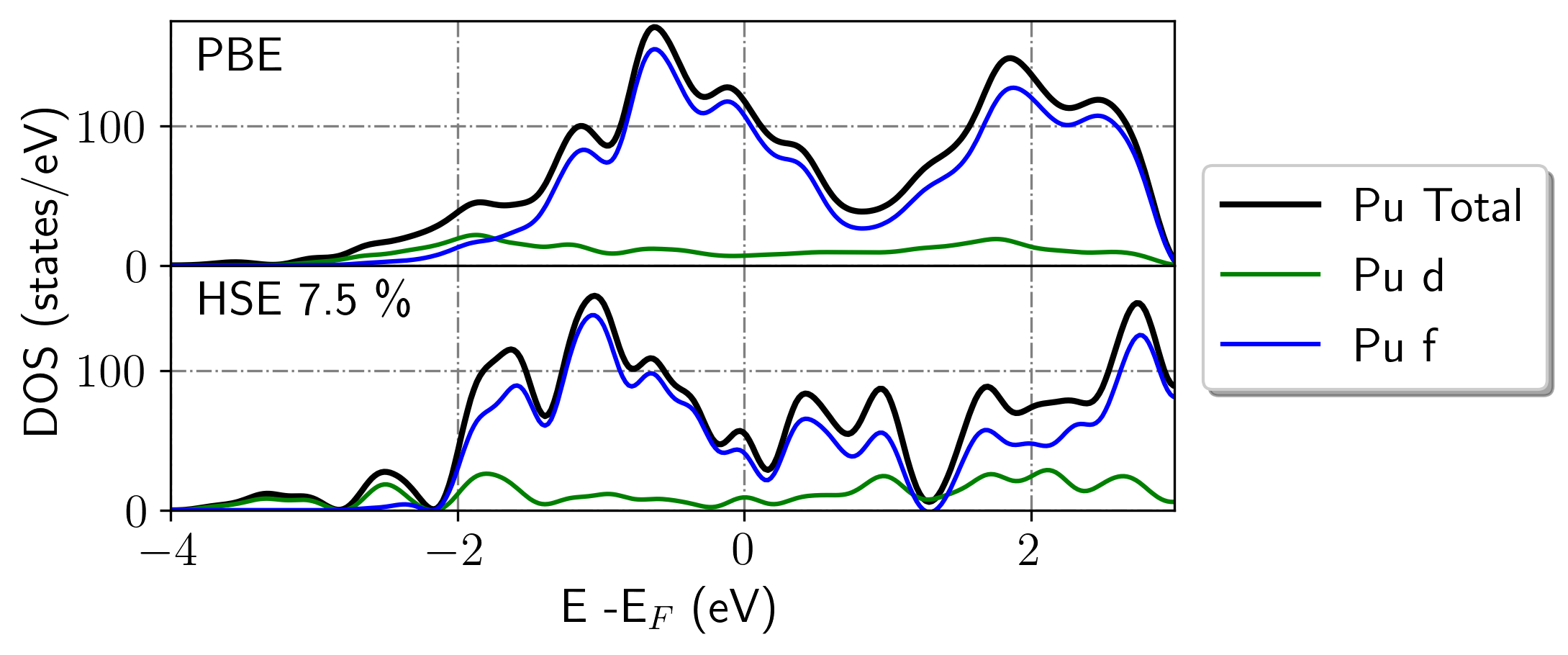}
\caption{Comparison between the density of states for PBE and HSE(7.5\%). The HSE result shows signatures of localization that are not present in PBE including peaked structures and the emergence of a pseudogap-like feature at the Fermi level.}
\label{fig:dos_comp}
\end{figure}

To distinguish between the bonding in HSE(7.5\%) and PBE, we compute the density of states (DOS) for each functional, Figure~\ref{fig:dos_comp}.
In our case, there are additional peaked structures in the HSE result, that indicate further localization in the f-band where the f-band centers are -1.9 eV for PBE and -2.2 eV for HSE(7.5\%). 
The narrow downshifted bands are indicative of correlation induced electron localization, which can also be seen with DFT+U.
Notably, HSE(7.5\%) suppresses the projected 5f DOS at the Fermi level, introducing a pseudogap-like feature that has been previously observed.\cite{wartenbe_2022}
This feature is a combination of a depressed f-orbital contribution and a peak in the d-orbital DOS.

\begin{figure}
\centering
\includegraphics[width=3in]{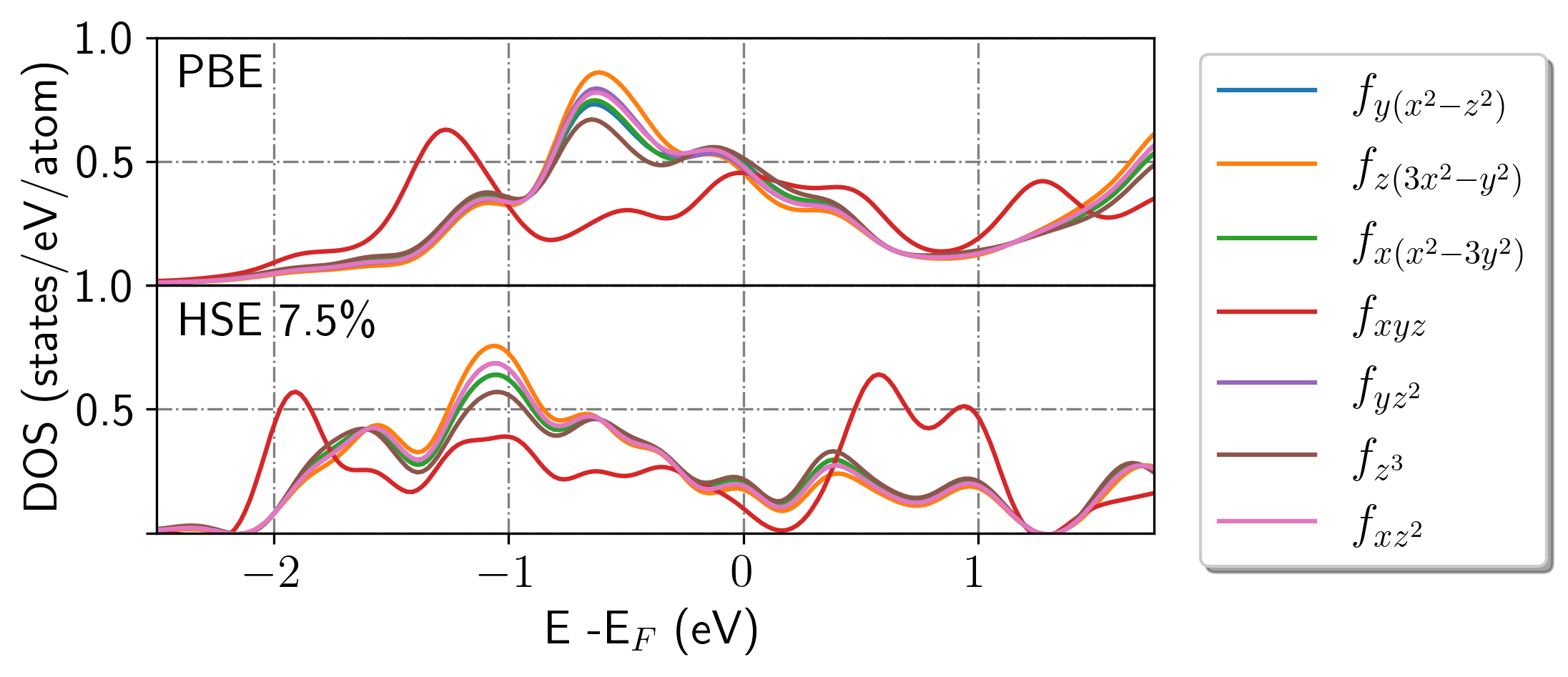}
\caption{The 5f DOS decomposed across l and m. Both functionals show six orbitals behave identically while one orbital (the xyz orbital) is more itinerant. However, all bands in HSE are decreased at the Fermi level creating a pseudogap-like feature.}
\label{fig:fdos_comp}
\end{figure}

Given that the f-orbitals are the largest contribution to the DOS, we study the bonding further by turning to the lm-decomposed DOS for the f-orbitals.
The f-orbital contributions to the DOS are shown for PBE and HSE(7.5\%) in Figure~\ref{fig:fdos_comp}.
The 5f bands change most between functionals and highlight 5f electron localization in $\delta$-Pu.
For both PBE and HSE(7.5\%), there are six nearly identical DOS curves and one distinct orbital.
Each orbital is shifted to lower energy, $\approx$ 0.5 eV, in HSE(7.5\%).
The $f_\text{xyz}$ orbital retains a distinct electronic structure in both theories, and is more diffuse than the other orbitals, indicating a higher degree of itinerant behavior.
This behavior is enabled by the symmetry of the orbital, $a_{2u}$, which lies in the higher energy $j=\frac{7}{2}$ manifold in an intermediate coupling regime.\cite{moore_2003}
Within the $j=\frac{7}{2}$ manifold, the states are split by the ligand field and the $a_{2u}$ states are preferentially split to lower energy due to the state's high spatial overlap with neighboring atoms.\cite{moore_2004,tobin_2005}
The change in the strength of the orbital overlaps indicate how the differences between PBE and HSE(7.5\%) precipitate.

Referring to Figure~\ref{fig:dos_comp}, we see that at -2 eV, the d-orbitals have a local maxima in their densities where there is a peak in the $f_{\text{xyz}}$ DOS.
This feature is more pronounced for HSE(7.5\%).
Therefore, with the hybrid functional, we are observing $\pi$-bonding that is more pronounced as well as a decreased $\sigma$-bonding due to the filling of both $\sigma$ and $\sigma^*$ states at -1.5 and -0.3 eV, respectively.
The DOS reveals an orbital selective state in $\delta$-Pu by shifting the $\pi^*$ states above the Fermi energy, allowing only the occupation of the $\pi$ states. 
These states are at -2 and 0.6 eV, respectively.
The screening and $E_{xx}$ interactions introduced in HSE(7.5\%) shift the 5f-states to create the pseudogap-like feature in the DOS.
The orbital selective state is highlighted in the real-space charge densities and the differences in bonding.
The total charge density and magnetization density differences between HSE and PBE are included in the Supplement.
The total charge density highlights the orbital selective behavior shown in the DOS with PBE having a more spherical charge density.
The magnetization density further highlights the high occupations of the localized orbitals.
The partial charge density, shown in the Supplement, corresponding to the localized feature at $\approx$ 1 eV, shows the localization of the six orbitals around Pu atoms in HSE. 

\begin{figure}
\centering
\includegraphics[width=3in]{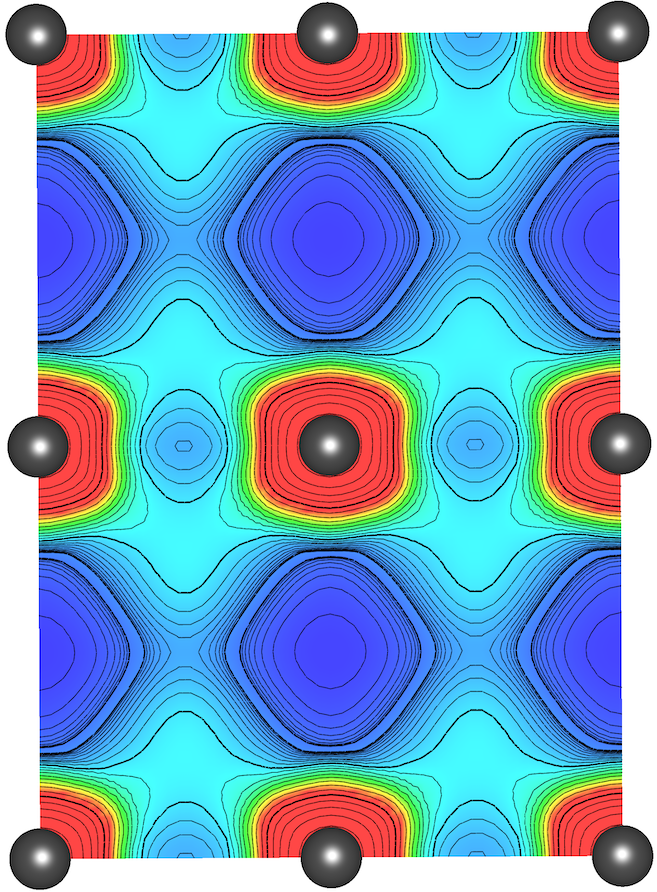}
\caption{The difference between the partial charges ($\rho_{HSE}-\rho_{PBE}$) in the (110)-plane. The partial charge is computed with the [-2 eV, -1.8 eV] energy window. This window corresponds to the observed bonding between $f_{xyz}$ and the d-orbitals in Figures~\ref{fig:dos_comp} and ~\ref{fig:fdos_comp}. Red indicates where HSE places more charge density and blue shows where $\rho_{HSE} \approx \rho_{PBE}$. While the HSE functional does show localization on the Pu sites, from the six localized f-orbitals, the bonding $f_{xyz}$ orbital is shown to participate in $\pi$-bonding by the arc features between Pu atoms.}
\label{fig:bonding}
\end{figure}

To understand how the differences in bonding soften the elastic constants HSE(7.5\%), we study the partial charge corresponding to the $f_{xyz}$.
The computed partial charge density differences for the bonding window, [-2 eV, -1.8 eV], are shown in Figure~\ref{fig:bonding}.
By looking at the (110)-plane, we highlight the bonding between nearest neighbors, as shown from left-to-right.
The blue areas show where $\rho_{HSE} \approx \rho_{PBE}$ and the red shows where HSE has the highest density with respect to PBE.
We observe increased $\pi$-bonding in HSE(7.5\%) relative to PBE, as observed in the DOS.
The minimum along the bond axis shows the reduced $\sigma$-bonding in HSE that was observed in the DOS due to $\sigma^*$ filling.
This arrangement confirms that there is one 5f electron participating in bonding in $\delta$-Pu and the correct description of this bonding is necessary for the accurate description of mechanical properties.
The increase in $\pi$-bonding, and the decrease in $\sigma$-bonding, work to soften the elastic constants.
By symmetry, $\pi$-bonding primarily softens $C'$.\cite{eberhart_1996}
The increases in $E_{xx}$ enable the orbital-selective behavior that increases the $\pi$-bonding and decreases the $\sigma$-bonding.
Including $E_{xx}$ leads to the orbital-selective behavior of the mixed-level model and captures the structural and mechanical properties of $\delta$-Pu, however, the mechanism for this orbital-selective behavior is still unclear.
\begin{figure}
\centering
\includegraphics[width=3in]{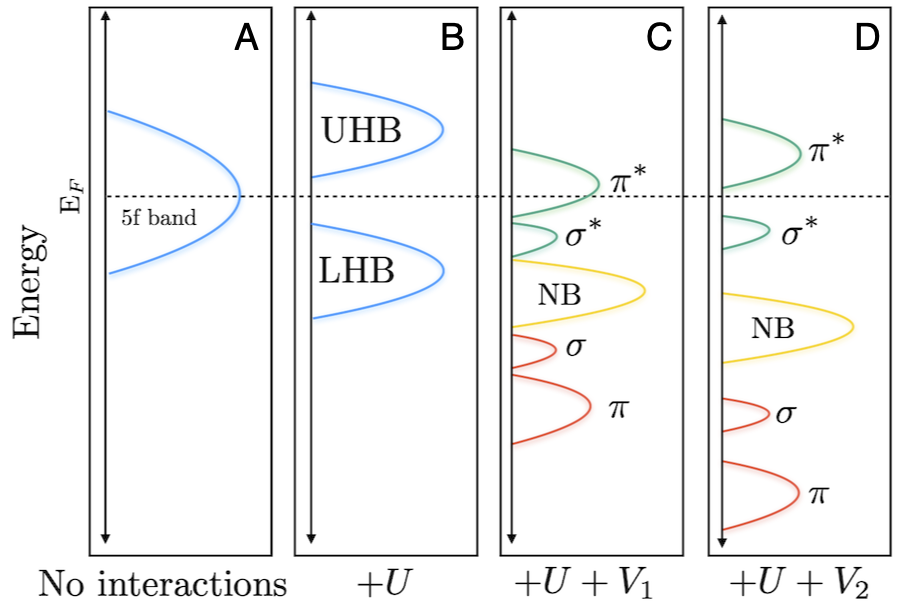}
\caption{A schematic of the 5f DOS as a function of the included interactions. On including the Hubbard U, the DOS splits into an upper (UHB) and lower Hubbard band (LHB). The inclusion of the intersite Anderson-like correlations through PBE ($V_1$) leads to a splitting of the lower Hubbard band into non-bonding (NB), $\sigma$, and $\pi$ bonding and antibonding pairs. As the hybridization increases ($V_2>V_1$) in the HSE case, the splitting of the bands increases leading to a completely unoccupied $\pi^*$, as we see in the case of the 5f DOS for $\delta$-Pu.}
\label{fig:schematic}
\end{figure}

The lattice expansion and softening introduced by $E_{xx}$ can be understood through two distinct mechanisms.
We can understand HSE(7.5\%) through analogy to DFT+U+J+V where V is a hybridization term.\cite{rubio_2020}
The expansion of the lattice is driven by on-site correlations (U,J), while $\pi$-bonding is driven by increases in the hybridization term. 
As the Hubbard U increases, comparing Figure~\ref{fig:schematic}a and ~\ref{fig:schematic}b, the lattice expands due to the localization of charge on Pu sites and the on-site interactions present in DFT+U+J stiffen the elastic constants when $a=a_0$.\cite{eberhart_1996}

The intersite interactions are necessary to simultaneously optimize the lattice parameter and the elastic constants, so we turn to the effect of the hybridization on the density of states.

The intersite correlations split the f-band into bonding and antibonding components for the $\sigma$ and $\pi$ bonding.\cite{greiner_2018}
This is shown in Figure~\ref{fig:schematic}c, where the introduction of the hybridization term, $V$, splits the lower Hubbard band into two sets of bonding and antibonding states.
We interpret the hybridization using an Anderson impurity model where the hybridization acts as a coupling between a Pu atom and the environment of Pu atoms.\cite{anderson_1961,newns_1969}
The hybridization characterizes the DOS in the Anderson model as,
\begin{equation}
  A(\omega) = \frac{V_{\pi,\sigma}}{(\omega-E)^2+V_{\pi,\sigma}^2},
\end{equation}
where $\omega$ is the energy of an electron relative to the impurity energy level, $E$, and we have denoted the different hybridizations for the $\pi$ and $\sigma$ states.
The PBE results are represented by the introduction of $V_1$, a small hybridization term. 
PBE and PBE+U have the same limited intersite hybridization and the localized $\sigma$-bonding and $\pi$-bonding states lie close to the main 5f band.
In the case of PBE, the 5f band has not been split into a upper Hubbard band (UHB) and lower Hubbard band (LHB) leaving the 5f band center close to the Fermi energy.
This results in unoccupied antibonding states, and, consequentially, a lattice contraction. 
For PBE+U, the LHB and UHB have formed, moving the 5f band center below the Fermi energy.
However, the limited hybridization leaves the antibonding states occupied, leading to a lattice contraction and stiffened elastic moduli.
On the other hand, HSE(7.5\%) captures the UHB and LHB while increasing the hybridization.
The increased hybridization increases the energy of the $\pi^*$ stateabove the Fermi energy while leaving the $\sigma^*$ states occupied.
The $\pi$ states are preferentially split by the ligand field effect on the $j=\frac{7}{2}$ states.
This leads to a $\pi$-bonding $a_{2u}$ and increases the $\pi^*$ above the Fermi energy effectively leaving one $\pi$-bonding 5f electron and 4 non-bonding electrons.
As a consequence, HSE simultaneously obtains $a_0$ and the experimental lattice parameters through the inclusion of the intersite correlations. 

By using the HSE functional's $E_{xx}$ to match the electronic screening, we found that the hybrid functional improves upon the description of elastic constants and electronic correlations in $\delta$-Pu. 
In addition, we use the DFT+U+J scheme to develop a model for the understanding of how HSE(7.5\%) accurately captures the ground-state properties of the system. 

We observed that the DOS computed with HSE(7.5\%) both increased localization and created a pseudogap-like feature.
By turning to the f-orbital DOS, we see that a pseudogap feature is created by shifting the f-orbitals lower in energy, localizing four electrons into six 5f orbitals.
The selective localization leaves one electron to engage in $\pi$-bonding by depopulating the $\pi^*$ states, indicating the importance of orbital selective physics in $\delta$-Pu.
The $\pi$-bonding is enhanced by the hybrid functional while the orbitals engaging in $\sigma$-bonding are localized, in that, the $\sigma$ and $\sigma^*$ states are both filled.
To understand the mechanism for the improvement on structural parameters, we use results from DFT+U+J.
The inclusion of a Hubbard U localizes the $\sigma$-bond causing an increase in the volume, and the same behavior is observed in HSE(7.5\%).
We observe, however, that U and J, can not in conjunction optimize the lattice constant and the elastic constants.
We then determine that a hybridization term drives the orbital-selective bonding that allows for the accurate determination of $\delta$-Pu's elastic properties. 

The authors would like to acknowledge the LDRD project number 20230042DR. 
This research used resources provided by the Los Alamos National Laboratory Institutional Computing Program, which is supported by the U.S. Department of Energy National Nuclear Security Administration under Contract No. 89233218CNA000001.

\bibliographystyle{unsrt}
\bibliography{hybrid_draft}
\section{Supplementary Information}
\appendix
\input{hybrid_supplement.tex}
\end{document}

%% file: hybrid_supplement.tex
\begin{figure}[!b]
\centering
\includegraphics[width=3in,height=3in]{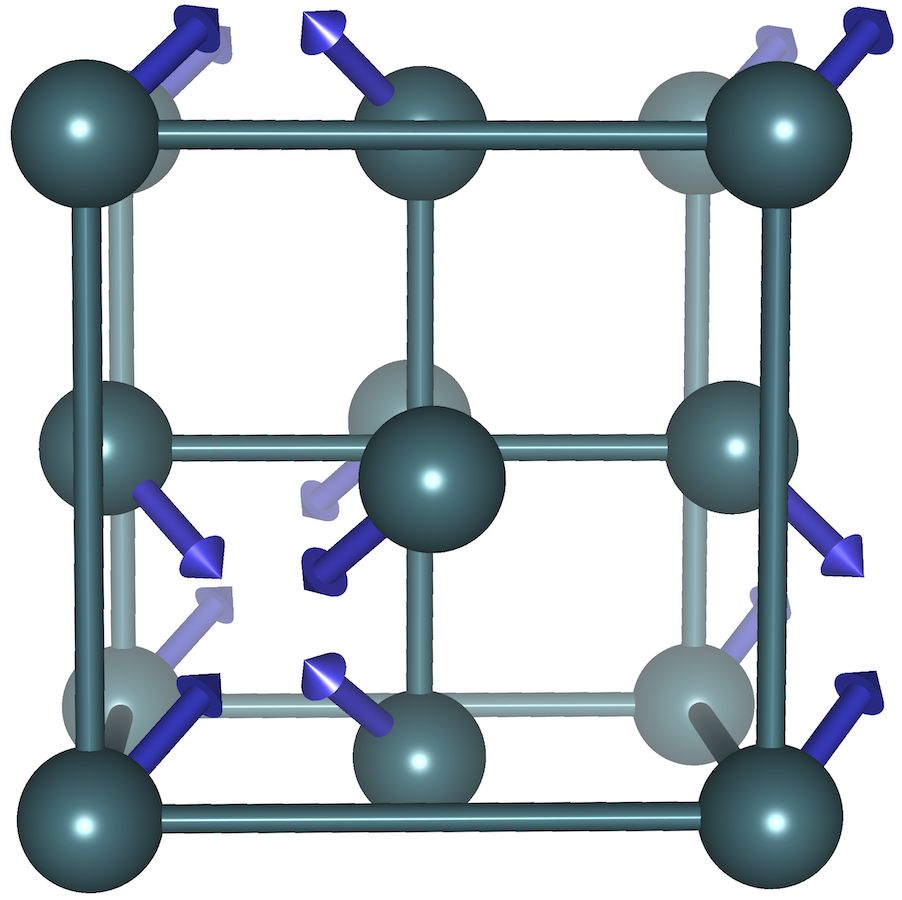}
\caption{The 3Q magnetic structure in $\delta$-Pu. The teal spheres are the Pu atoms in the face-centered cubic arrangement, and the blue arrows represent the direction of the magnetic moments on the Pu sites. For a set of four Pu atoms, the magnetic moments point toward the tetrahedral interstitial such that each structurally equivalent bond is equivalent in its bonding character. Visualization was done in VESTA.}
\label{fig:3q}
\end{figure}

\section{Tuning Exact Exchange in HSE to the Experimental Lattice Parameter}

\begin{figure*}
\centering
\includegraphics[width=3in]{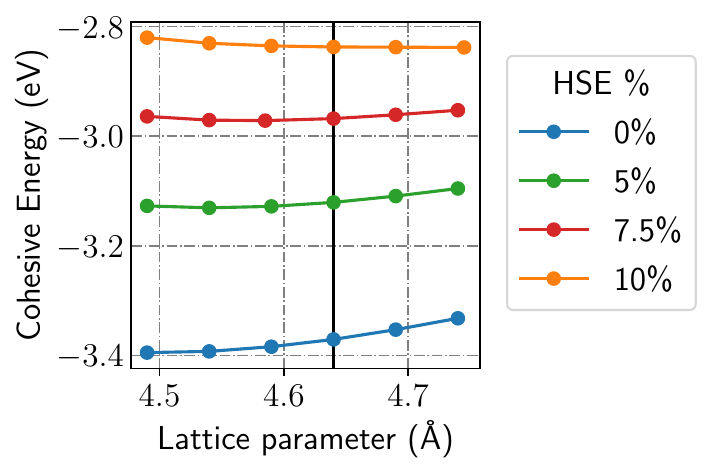}
\caption{The cohesive energy plotted against the lattice parameter for a set of functionals: PBE and HSE(a\%). The vertical line shows the experimental lattice constant. PBE has been omitted from this plot for clarity.}
\label{fig:energy_vol}
\end{figure*}

\section{3Q Magnetic Moments as a function of Volume}

\begin{figure}[!b]
\centering
\includegraphics[width=3in]{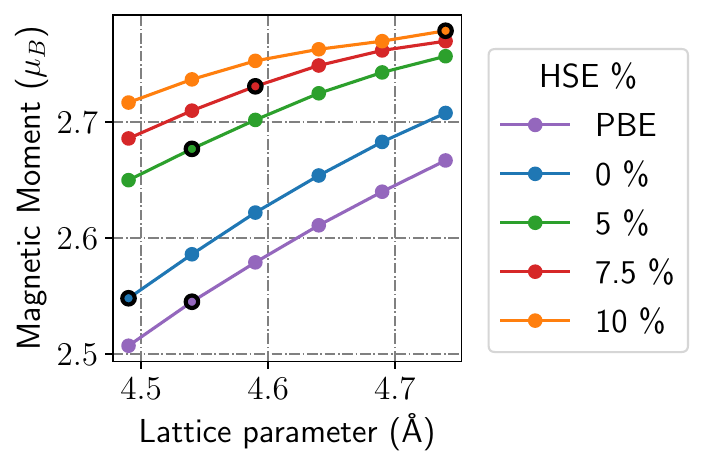}
\caption{The 3Q magnetic moment as a function of the lattice parameter. The black circles show where each functional is in its respective equilibrium lattice parameter.}
\label{fig:magmom_vol}
\end{figure}

We study the evolution of the 3Q magnetic state as a function of the DFT functional and lattice parameter in Figure~\ref{fig:magmom_vol}.
The larger moments are indicative of more atomic-like 5f electron behavior that drives the system to larger cell sizes.
The magnetic moments increase as the lattice parameter increases for each functional, and it appears to be converging to a fixed value at large lattice constant.
We observe that the PBE magnetic moment is smallest and the HSE magnetic moment increases with the exact exchange mixing.
In each case, the spin magnetic moment is largely cancelled out with the orbital magnetic moment as observed in the literature for collinear antiferromagnetic states.
This makes the magnetic state used to model $\delta$-Pu reflect the paramagnetic state by making the apparent magnetic moment on each site smaller.

\section{Charge Density and Magnetization Density}

\begin{figure*}
\centering
\includegraphics[width=4in]{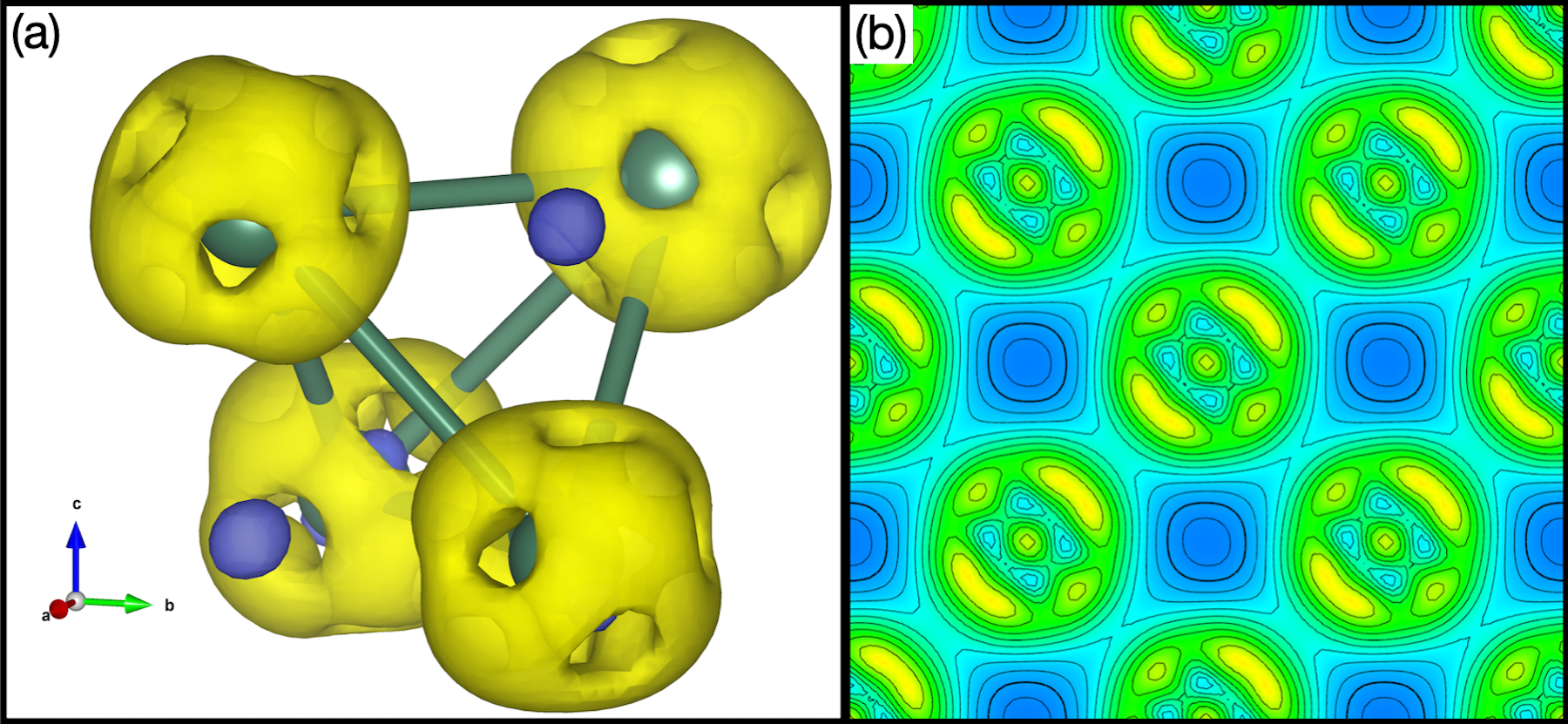}
\caption{Plot of the difference in the real space charge density, PBE-HSE. The yellow areas indicate where PBE places more charge density and blue is where HSE places more charge density. }
\label{fig:charge}
\end{figure*}

The difference between the charge density calculated by PBE and HSE is shown in Figure~\ref{fig:charge}.
The PBE-HSE difference highlights the distribution in each functional by showing where PBE places more charge density in yellow and where HSE(7.5\%) places more charge density in blue.
Figure~\ref{fig:charge}b shows that HSE places moves charge in the $O_h$ holes while PBE distributes the charge spherically around the Pu atom. 
The PBE charge density shows a spherical distribution indicating  that the charge in the f-orbitals is more evenly distributed across the f-orbitals.
The inclusion of exact exchange in HSE(7.5\%) causes the orbitals to localize and form larger magnetic moments. 
The charge density in the 111 direction, toward the tetrahedral  interstitial, is the same between HSE and PBE, demonstrating a preference for particular orbitals in the reorganization of charge.

\begin{figure*}
\centering
\includegraphics[width=4in]{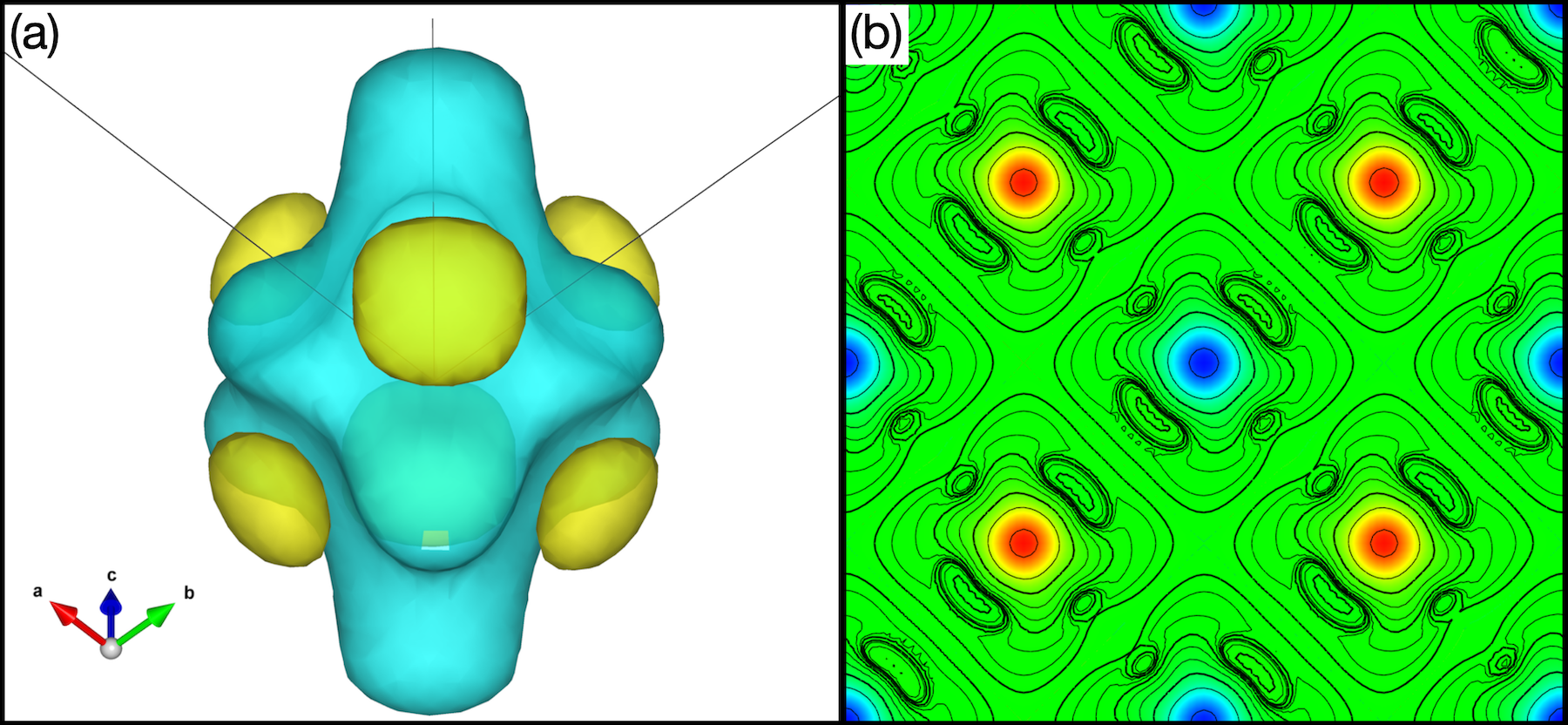}
\caption{Plot of the difference in the magnetization density. The yellow areas indicate where PBE places more magnetization density and blue is where HSE places more magnetization density.}
\label{fig:magnetization}
\end{figure*}

The difference of the magnetization densities between the methods is shown in Figure~\ref{fig:magnetization}.
The HSE spin density highlights the orbital selective nature of $\delta$-Pu.
The HSE(7.5\%) functional places the magnetization density around the Pu atom in a subset of the f-orbitals while the PBE spin density becomes more spherically symmetric, as expected from the charge density.
The HSE magnetization density points toward the tetrahedral interstitial site moreso than the PBE density, indicating a selective occupation of orbitals.
The pattern of the magnetization density indicates that the magnetization density in HSE is localizing in a subset of the f-orbitals, creating the larger magnetic moment. 
These results are sensible given the usage of hybrid functionals and DFT+U to describe magnetic systems and $\delta$-Pu.

\subsection{Signatures of Localization from the Partial Charge Differences}
\begin{figure}
\centering
\includegraphics[width=3in]{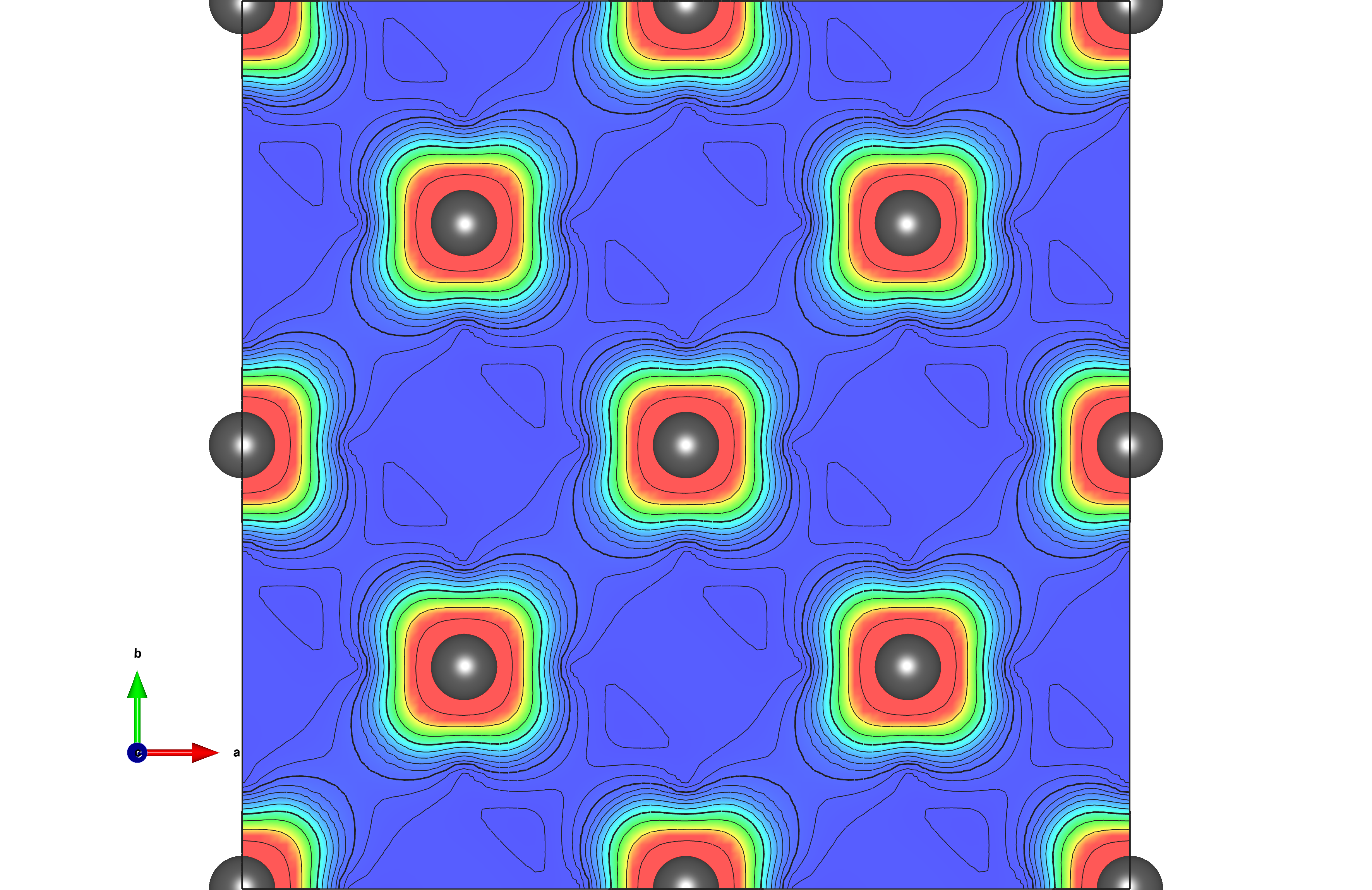}
\caption{The difference between the partial charges ($\rho_{HSE}-\rho_{PBE}$) in the (010)-plane. The partial charges computed correspond to the localization observed in the HSE DOS. Red indicates where HSE places more charge density and blue shows where $\rho_{HSE} \approx \rho_{PBE}$. This plot highlights the increased localization on the Pu atoms when using the HSE functional.}
\label{fig:localization}
\end{figure}

To study the localization behavior in HSE, we computed the partial charge corresponding to the peaks in the total DOS.
The partial charges are computed using the [-1.1 eV, -0.9 eV] window of HSE and the [-0.9, -0.7 eV ] window of PBE. 
These windows correspond to the localization in HSE and take into account the overall shift in the DOS.
The differences between the partial charge densities are shown in Figure~\ref{fig:localization}.
We observe that HSE places more charge density around the Pu atoms, and there is no difference in the bonding between theories.
The six f-electrons responsible for the density of states at this energy are strongly localized in HSE, as confirmed by the DOS.

We performed additional studies on the SCAN and OFR2 functionals, as shown in Table~\ref{table:elastic}.
This table also demonstrated the effect of constraining the magnetic moments on the elastic constants.
Calculations were performed at 12x12x12 k-points using the VASP internal routines.
Elastic constants were averaged.

\begin{table*}
    \centering
	\begin{tabular}{l|c|c|c|c|c|c|}
		Functional &a$_0$ ($\text{\AA}$) & C$_{11}$ (GPa) & C$_{12}$ (GPa) & C$_{44}$ (GPa) & Constrained?  \\
		\hline
		PBE(Eq.) & 4.54  & 60.5 & 47.0 & 40.3 & No \\
        PBE(Eq.) & 4.54  & 61.7 & 57.6  & 38.0 & Yes \\
        PBE(Exp.)& 4.64 & 50.1 & 34.0  & 34.8  & No\\
		SCAN & 4.64  & 57.6 & 19.6 & 41.2 & No \\
		OFR2 & 4.64  & 46.7 & 26.4 & 31.3 & No \\
		U(0 eV),J(1.1 eV) & 4.62  & 26.8 & 43.6 & 39.3 & No \\
		U(0 eV),J(1.2 eV) & 4.64  & 28.0 & 39.8 & 38.2 & No \\
		U(1 eV),J(0 eV) & 4.60  & 66.3 & 30.8 & 46.6 & No \\
		U(1 eV),J(0.3 eV) & 4.64  & 53.0 & 33.6 & 45.1 & No \\
		U(1 eV),J(0.5 eV) & 4.67  & 57.4 & 33.2 & 40.9 & No \\
	\end{tabular} 
    \caption{The elastic constant results for SCAN, OFR2, DFT+U+J in the Liechtenstein approach, and PBE with and without magnetic moment constraints. SCAN shows a significantly stiffened C$_{11}$. The OFR2 result mirrors the PBE results for the experimental lattice. The effect of the magnetization constraint is significant for C$_{12}$, stiffening it by $\approx$20 percent while leaving the other elastic constants within errors of the method. The 3Q magnetic state is very particular to the face-centered cubic arrangement in $\delta$-Pu. Breaking this symmetry leads to a breaking of the bond-equivalent magnetism, leading to a stiffening in trying to maintain the magnetic structure.}
    \label{table:elastic}
\end{table*}

%% file: hybrid_draft.bbl
\begin{thebibliography}{10}

\bibitem{soderlind_2019}
A.~Landa Per~S{\"o}derlind and B.~Sadigh.
\newblock Density-functional theory for plutonium.
\newblock {\em Advances in Physics}, 68(1):1--47, 2019.

\bibitem{albers_2001}
RC~Albers.
\newblock An expanding view of plutonium.
\newblock {\em Nature}, 410(6830):759--761, 2001.

\bibitem{hecker_2000}
Siegfried~S. Hecker.
\newblock Plutonium and its alloys from atoms to microstructure.
\newblock 2000.

\bibitem{huang_2020}
Li~Huang and Haiyan Lu.
\newblock Nature of the $5f$ electronic structure of plutonium.
\newblock {\em Phys. Rev. B}, 101:125123, Mar 2020.

\bibitem{penicaud_1997}
M~P{\'e}nicaud.
\newblock Electron localization in the series of actinide metals. the cases of
  $\delta$-pu and es.
\newblock {\em Journal of Physics: Condensed Matter}, 9(30):6341, 1997.

\bibitem{eriksson_1999}
Olle Eriksson, J.~D Becker, A.~V Balatsky, and J.~M Wills.
\newblock Novel electronic configuration in $\delta$-pu.
\newblock {\em Journal of Alloys and Compounds}, 287(1):1--5, 1999.

\bibitem{soderlind_2003}
Per S{\"o}derlind and Alex Landa.
\newblock Simple model for localization in $\delta$-pu.
\newblock {\em Modelling and Simulation in Materials Science and Engineering},
  11(6):851, 2003.

\bibitem{svane_2007}
A.~Svane, L.~Petit, Z.~Szotek, and W.~M. Temmerman.
\newblock Self-interaction-corrected local spin density theory of $5f$ electron
  localization in actinides.
\newblock {\em Phys. Rev. B}, 76:115116, Sep 2007.

\bibitem{soderlind_2010}
Per S\"oderlind, Alex Landa, J.~E. Klepeis, Y.~Suzuki, and A.~Migliori.
\newblock Elastic properties of pu metal and pu-ga alloys.
\newblock {\em Phys. Rev. B}, 81:224110, Jun 2010.

\bibitem{xu_2008}
Ruqing Xu, Joe Wong, Paul Zschack, Hawoong Hong, and Tai-C. Chiang.
\newblock Soft phonons in $\delta$-phase plutonium near the $\delta$-$\alpha$'
  transition.
\newblock {\em Europhysics Letters}, 82(2):26001, mar 2008.

\bibitem{boring_2000}
A.~Michael Boring and James~L. Smith.
\newblock Plutonium condensed-matter physics a survey of theory and experiment.
\newblock 2000.

\bibitem{clark_2019}
David~L. Clark, David~A. Gleeson, and Robert J.~Hanrahan Jr., editors.
\newblock {\em Plutonium Handbook}, volume~2.
\newblock American Nuclear Society, 2nd edition edition, 2019.

\bibitem{amadon_2016}
Bernard Amadon.
\newblock First-principles dft+dmft calculations of structural properties of
  actinides: Role of hund's exchange, spin-orbit coupling, and crystal
  structure.
\newblock {\em Phys. Rev. B}, 94:115148, Sep 2016.

\bibitem{soderlind_1997}
Per S\"oderlind, J.~M. Wills, B.~Johansson, and O.~Eriksson.
\newblock Structural properties of plutonium from first-principles theory.
\newblock {\em Phys. Rev. B}, 55:1997--2004, Jan 1997.

\bibitem{tutchton_2020}
Roxanne~M. Tutchton, Wei-ting Chiu, R.~C. Albers, G.~Kotliar, and Jian-Xin Zhu.
\newblock Electronic correlation induced expansion of fermi pockets in
  $\ensuremath{\delta}$-plutonium.
\newblock {\em Phys. Rev. B}, 101:245156, Jun 2020.

\bibitem{kutepov_2023}
A~L Kutepov, J~G Tobin, S-W Yu, B~W Chung, and P~Roussel.
\newblock New insights into the electronic structure of $\alpha$-u and
  $\delta$-pu.
\newblock {\em Journal of Physics: Condensed Matter}, 36(4):045601, oct 2023.

\bibitem{shick_2005}
A.~B. Shick, V.~Drchal, and L.~Havela.
\newblock Coulomb-u and magnetic-moment collapse in $\delta$-pu.
\newblock {\em Europhysics Letters}, 69(4):588, 2005.

\bibitem{raymond_2009}
R.~Atta-Fynn and A.~K. Ray.
\newblock Does hybrid density functional theory predict a non-magnetic ground
  state for $\delta$-pu?
\newblock {\em Europhysics Letters}, 85(2):27008, feb 2009.

\bibitem{shimazaki_2015}
Tomomi Shimazaki and Takahito Nakajima.
\newblock {Theoretical study of a screened Hartree--Fock exchange potential
  using position-dependent atomic dielectric constants}.
\newblock {\em The Journal of Chemical Physics}, 142(7):074109, 02 2015.

\bibitem{skone_2014}
Jonathan~H. Skone, Marco Govoni, and Giulia Galli.
\newblock Self-consistent hybrid functional for condensed systems.
\newblock {\em Phys. Rev. B}, 89:195112, May 2014.

\bibitem{heyd_2003}
Jochen Heyd, Gustavo~E. Scuseria, and Matthias Ernzerhof.
\newblock {Hybrid functionals based on a screened Coulomb potential}.
\newblock {\em The Journal of Chemical Physics}, 118(18):8207--8215, 04 2003.

\bibitem{blochl_1994}
P.~E. Bl\"ochl.
\newblock Projector augmented-wave method.
\newblock {\em Phys. Rev. B}, 50:17953--17979, Dec 1994.

\bibitem{kresse_1996}
G.~Kresse and J.~Furthm\"uller.
\newblock Efficient iterative schemes for ab initio total-energy calculations
  using a plane-wave basis set.
\newblock {\em Phys. Rev. B}, 54:11169--11186, Oct 1996.

\bibitem{kresse_1999}
G.~Kresse and D.~Joubert.
\newblock From ultrasoft pseudopotentials to the projector augmented-wave
  method.
\newblock {\em Phys. Rev. B}, 59:1758--1775, Jan 1999.

\bibitem{methfessel_1989}
M.~Methfessel and A.~T. Paxton.
\newblock High-precision sampling for brillouin-zone integration in metals.
\newblock {\em Phys. Rev. B}, 40:3616--3621, Aug 1989.

\bibitem{monkhorst_1976}
Hendrik~J. Monkhorst and James~D. Pack.
\newblock Special points for brillouin-zone integrations.
\newblock {\em Phys. Rev. B}, 13:5188--5192, Jun 1976.

\bibitem{rudin_2022}
Sven~P. Rudin.
\newblock Symmetry-correct bonding in density functional theory calculations
  for delta phase pu.
\newblock {\em Journal of Nuclear Materials}, 570:153954, 2022.

\bibitem{soderlind_2023}
Per S\"oderlind, A.~Landa, L.~H. Yang, B.~Sadigh, and Sven~P. Rudin.
\newblock Lattice dynamics and thermodynamics for
  $\ensuremath{\delta}$-plutonium from density functional theory.
\newblock {\em Phys. Rev. B}, 108:104112, Sep 2023.

\bibitem{lashley_2005}
J.~C. Lashley, A.~Lawson, R.~J. McQueeney, and G.~H. Lander.
\newblock Absence of magnetic moments in plutonium.
\newblock {\em Phys. Rev. B}, 72:054416, Aug 2005.

\bibitem{janoschek_2015}
Marc Janoschek, Pinaki Das, Bismayan Chakrabarti, Douglas~L. Abernathy, Mark~D.
  Lumsden, John~M. Lawrence, Joe~D. Thompson, Gerard~H. Lander, Jeremy~N.
  Mitchell, Scott Richmond, Mike Ramos, Frans Trouw, Jian-Xin Zhu, Kristjan
  Haule, Gabriel Kotliar, and Eric~D. Bauer.
\newblock The valence-fluctuating ground state of plutonium.
\newblock {\em Science Advances}, 1(6):e1500188, 2015.

\bibitem{ellinger_1956}
F.~H. Ellinger.
\newblock Crystal structure of delta-prime plutonium and the thermal expansion
  characteristics of delta, delta-prime, and epsilon plutonium.
\newblock {\em JOM}, 8(10):1256--1259, 1956.

\bibitem{bercegeay_2005}
C.~Bercegeay and S.~Bernard.
\newblock First-principles equations of state and elastic properties of seven
  metals.
\newblock {\em Phys. Rev. B}, 72:214101, Dec 2005.

\bibitem{perdew_1996}
John~P. Perdew, Kieron Burke, and Matthias Ernzerhof.
\newblock Generalized gradient approximation made simple.
\newblock {\em Phys. Rev. Lett.}, 77:3865--3868, Oct 1996.

\bibitem{kaplan_2022}
Aaron~D. Kaplan and John~P. Perdew.
\newblock Laplacian-level meta-generalized gradient approximation for solid and
  liquid metals.
\newblock {\em Phys. Rev. Mater.}, 6:083803, Aug 2022.

\bibitem{dudarev_1998}
S.~L. Dudarev, G.~A. Botton, S.~Y. Savrasov, C.~J. Humphreys, and A.~P. Sutton.
\newblock Electron-energy-loss spectra and the structural stability of nickel
  oxide: An lsda+u study.
\newblock {\em Phys. Rev. B}, 57:1505--1509, Jan 1998.

\bibitem{liechtenstein_1995}
A.~I. Liechtenstein, V.~I. Anisimov, and J.~Zaanen.
\newblock Density-functional theory and strong interactions: Orbital ordering
  in mott-hubbard insulators.
\newblock {\em Phys. Rev. B}, 52:R5467--R5470, Aug 1995.

\bibitem{vega_2018}
Lorena Vega, Judit Ruvireta, Francesc Vi{\~n}es, and Francesc Illas.
\newblock Jacob's ladder as sketched by escher: Assessing the performance of
  broadly used density functionals on transition metal surface properties.
\newblock {\em Journal of Chemical Theory and Computation}, 14(1):395--403, 01
  2018.

\bibitem{janthon_2013}
Patanachai Janthon, Sergey~M. Kozlov, Francesc Vi{\~n}es, Jumras Limtrakul, and
  Francesc Illas.
\newblock Establishing the accuracy of broadly used density functionals in
  describing bulk properties of transition metals.
\newblock {\em Journal of Chemical Theory and Computation}, 9(3):1631--1640, 03
  2013.

\bibitem{gorni_2021}
Tommaso Gorni, Pablo Villar~Arribi, Michele Casula, and Luca de' Medici.
\newblock Accurate modeling of fese with screened fock exchange and hund metal
  correlations.
\newblock {\em Phys. Rev. B}, 104:014507, Jul 2021.

\bibitem{wartenbe_2022}
Mark Wartenbe, Paul~H. Tobash, John Singleton, Laurel~E. Winter, Scott
  Richmond, and Neil Harrison.
\newblock Pseudogap in elemental plutonium.
\newblock {\em Phys. Rev. B}, 105:L041107, Jan 2022.

\bibitem{moore_2003}
K~T Moore, M~A Wall, A~J Schwartz, B~W Chung, D~K Shuh, R~K Schulze, and J~G
  Tobin.
\newblock Failure of russell-saunders coupling in the 5f states of plutonium.
\newblock {\em Phys Rev Lett}, 90(19):196404, May 2003.

\bibitem{moore_2004}
A.~J. Schwartz B. W. Chung S. A. Morton J. G. Tobin S. Lazar F. D. Tichelaar H.
  W. Zandbergen P.~S{\"o}derlind K.~T.~Moore, M. A.~Wall and G.~van~der Laan.
\newblock Electron-energy-loss spectroscopy and x-ray absorption spectroscopy
  as complementary probes for complex f-electron metals: cerium and plutonium.
\newblock {\em Philosophical Magazine}, 84(10):1039--1056, 2004.

\bibitem{tobin_2005}
J.~G. Tobin, K.~T. Moore, B.~W. Chung, M.~A. Wall, A.~J. Schwartz, G.~van~der
  Laan, and A.~L. Kutepov.
\newblock Competition between delocalization and spin-orbit splitting in the
  actinide $5f$ states.
\newblock {\em Phys. Rev. B}, 72:085109, Aug 2005.

\bibitem{eberhart_1996}
M.E. Eberhart.
\newblock The metallic bond: Elastic properties.
\newblock {\em Acta Materialia}, 44(6):2495--2504, 1996.

\bibitem{rubio_2020}
Nicolas Tancogne-Dejean and Angel Rubio.
\newblock Parameter-free hybridlike functional based on an extended hubbard
  model: $\mathrm{DFT}+u+v$.
\newblock {\em Phys. Rev. B}, 102:155117, Oct 2020.

\bibitem{greiner_2018}
M.~T. Greiner, T.~E. Jones, S.~Beeg, L.~Zwiener, M.~Scherzer, F.~Girgsdies,
  S.~Piccinin, M.~Armbr{\"u}ster, A.~Knop-Gericke, and R.~Schl{\"o}gl.
\newblock Free-atom-like d states in single-atom alloy catalysts.
\newblock {\em Nature Chemistry}, 10(10):1008--1015, 2018.

\bibitem{anderson_1961}
P.~W. Anderson.
\newblock Localized magnetic states in metals.
\newblock {\em Phys. Rev.}, 124:41--53, Oct 1961.

\bibitem{newns_1969}
D.~M. NEWNS.
\newblock Self-consistent model of hydrogen chemisorption.
\newblock {\em Phys. Rev.}, 178:1123--1135, Feb 1969.

\end{thebibliography}
